\documentclass[aps,preprint,preprintnumbers,showpacs,%
superscriptaddress,nofootinbib]{revtex4}
\usepackage{amsmath}
\usepackage{graphicx} 
\usepackage{amsmath,amssymb,epsfig,bm}

\newcommand{\p}{\chi} 

\def\a{\alpha}

\def\p{\pi}                
\def\t{\tau}

\def\t6 {T_\mt{D6}}

\newcommand{\N}{{\cal N}} 

\newcommand{\mt}[1]{\textrm{\tiny #1}}

\def\ofo{ { {}_2 \! F_1 }}

\begin{document}

\title{Zero Sound from Holography}
\author{A.~Karch}
\affiliation{Department of Physics, University of Washington, Seattle, 
WA 98195-1560, USA}
\author{D.\,T.~Son}
\affiliation{Institute for Nuclear Theory, University of Washington,
Seattle, WA 98195-1550, USA}
\author{A.\,O.~Starinets}
\affiliation{School of Natural Sciences, Institute for Advanced Study,
Einstein Drive, Princeton, NJ 08540, USA}
\affiliation{School of Physics and Astronomy, University of Southampton, 
Highfield, Southampton SO17 1BJ UK}
\date{June 2008}

%
%
%



\begin{abstract}
Quantum liquids are characterized by the distinctive properties such as 
the low temperature behavior of  heat capacity
and the spectrum of low-energy quasiparticle excitations.
In particular, at  
low temperature, Fermi liquids exhibit the
zero sound, predicted by L.~D.~Landau in 1957 and subsequently
observed in liquid He-3. 
In this paper, we ask a question whether such a characteristic behavior 
is present 
in theories with holographically dual description.
We consider a class of gauge theories with fundamental matter fields 
whose holographic dual 
in the appropriate limit is
given in terms of the Dirac-Born-Infeld action in $AdS_{p+1}$ space. 
An example of such a system is the ${\cal N}=4$ $SU(N_c)$ supersymmetric
Yang-Mills theory with $N_{\!f}$  massless  
${\cal N}=2$ hypermultiplets at strong coupling, 
finite baryon number density, and low temperature.
We find that these systems  exhibit a zero sound mode 
despite 
having a non-Fermi liquid type behavior of the specific heat.
These properties suggest that holography identifies a new type 
of quantum liquids.
\end{abstract}

\maketitle



\pagestyle{plain}
\emph{Introduction.}---Gauge/gravity duality 
\cite{Maldacena:1997re, Gubser:1998bc, Witten:1998qj} 
has become a useful tool for investigating
strongly coupled field theories.  In the class of models where this
tool can be applied, the strong coupling limit of the field theory is
mapped into the weak-coupling, classical limit of a gravity theory,
which can be studied either analytically or with minimal computer
power.  For example, a cousin of QCD---the ${\cal N}=4$ supersymmetric
Yang-Mills (SYM) theory---has been studied using this method.  Such
studies have pointed to a universal value of the viscosity/entropy
density ratio in a wide class of strongly coupled theories 
(for a review, see \cite{Son:2007vk}).  
Somewhat surprisingly, the viscosity/entropy density ratio of the quark-gluon
plasma created at the Relativistic Heavy Ion Collider seems to be
close to this value, indicating that gauge/gravity duality may be
useful for studies of QCD.

Here we would like to see what the gauge/gravity duality has to say about
strongly coupled quantum liquids.  By quantum liquids we mean 
translationally invariant systems at zero (or low) temperature and at
finite density.  Given the important role that quantum liquids play in
physics, it is natural to ask whether the
newly developed technique of gauge/gravity duality can give us any
insights into their behavior.

The cornerstones of our understanding of quantum liquids are two
phenomenological theories.  These are Landau's Fermi liquid theory 
\cite{Landau:1956, Landau:1957, LL, Abrikosov, PinesNozieres} 
and the theory of quantum Bose liquids \cite{LL, Abrikosov}.  
These two theories describe two
distinct behaviors of a quantum liquid at low momenta and
temperatures.  In a Bose liquid, the only low-energy elementary
excitation is the superfluid phonon with a linear dispersion.  This
leads to a $T^3$ behavior of the specific heat at low temperatures.
The Fermi liquid has a richer spectrum of elementary excitations,
consisting of fermionic quasiparticles and a bosonic branch, which
contains, in particular, the zero sound.  The fermions dominate the
specific heat, which scales as $T$ at low $T$.

In this paper, we found, through the gauge/gravity duality, a new type
of quantum liquid.  The quantum liquid we consider has a $T^6$
behavior ($\sim T^{2p}$ in $p$ spatial dimensions) of the specific heat 
at low temperature.  Despite the non-Fermi liquid behavior 
of the specific heat, 
the system supports a sound mode at zero temperature, 
which we will call ``zero sound.''
The mode is almost
identical to the zero sound in Fermi liquids: not only the real part
of its dispersion curve is linear in momentum ($\omega=v\, q$), but the
imaginary part has the same $q^2$ dependence predicted by Landau a
long time ago for quantum attenuation of the zero sound.  The
difference is that in our case the zero sound velocity coincides with
the first sound velocity, while in the case of a Fermi liquid the two
velocities are not equal to each other.

In Fermi liquids, zero sound is a collective excitation involving
fermions near the Fermi surface.  It was predicted by Landau
\cite{Landau:1957} and experimentally observed in liquid Helium-3.  In
a weakly interacting Fermi gas, the zero sound velocity is close to
the Fermi velocity $v_F^{\phantom 1}$.  
The first sound, which is a hydrodynamic
mode that exists at finite temperatures and wavelengths larger than
the mean free path, has velocity $v_F^{\phantom 1}/\sqrt3$ at weak coupling.  
The quantum
attenuation (i.e. damping at zero temperature) of zero sound was first
considered by Landau, who showed that the imaginary part of the zero
sound energy scales as $q^2$ where $q$ is the momentum.  The
experimental situation with the measurement of the zero sound quantum
attenuation is summarized in ~\cite{He3-book}.

A specific example considered in this paper is the ${\cal N}=4$ SU($N_c$) 
supersymmetric Yang-Mills (SYM) theory with
$N_{\!f}$ massless ${\cal N}=2$ hypermultiplet fields.  This
theory has been suggested as a model which approximates QCD better 
than the theory without fundamental quarks.  
A string-theoretic description of this system is given by a low-energy 
limit of the D3/D7 brane
configuration.
The theory has been studied at finite temperature and
density using the gauge/gravity duality  \cite{Kobayashi:2006sb, Karch:2007pd, 
Karch:2007br,  Nakamura:2007nx, Ghoroku:2007re, 
Mateos:2007vc,  Erdmenger:2007ja, Matsuura:2007zx, Mas:2008jz}.  
Nevertheless, two striking aspects of this and similar systems 
(characterized by the Dirac-Born-Infeld action in 
Anti-de-Sitter space)---the
unusual behavior of the low-temperature specific heat and the
existence of the zero sound---have so far eluded attention. 
Their description is the main purpose and the main result of the paper.

\emph{Preliminaries.}---The ${\cal N}=4$ SYM theory contains fields 
in the adjoint 
representation of the gauge group only.  
Fields in the fundamental representation can be introduced by 
using the following construction \cite{Karch:2002sh}. 
In type IIB string theory, one considers a system of $N_c$ D3-branes 
and $N_{\!f}$ D7-branes aligned in flat ten-dimensional space as
\begin{equation}
\begin{array}{ccccccccccc}
   & x_0 & x_1 & x_2 & x_3 & x_4 & x_5 & x_6 & x_7 & x_8 & x_9\\
\mbox{D3} & \times & \times & \times & \times & & &  &  & & \\
\mbox{D7} & \times & \times & \times & \times & \times  & \times
& \times & \times &  &   \\
\end{array}
\end{equation}
In the limit of large number of colors ($N_c \gg 1$) 
and large 't Hooft coupling ($g^2_{\rm YM} N_c \gg 1$), 
the D3-branes are replaced by the 
near-horizon AdS$_5\times\textrm{S}^5$ geometry \cite{Maldacena:1997re}, 
while the $N_{\!f}$ D7-branes can be treated as  probes embedded into 
this geometry 
as long as $N_{\!f}/N_c \ll 1$, i.e. as long as their backreaction 
on the geometry can be neglected \cite{Karch:2002sh}.

The standard form of the near-horizon D3 brane metric is 
\begin{equation}
\label{metric}
ds^2 = \frac{r^2}{R^2}\, \eta_{\mu \nu}dx^{\mu} dx^{\nu} + \frac{R^2}{r^2}\, 
 \left( dr^2 + r^2 d\Omega_5^2 \right),
\end{equation}
where $\eta_{\mu \nu} = \textrm{diag} (-1,1,1,1)$, 
$R$ is the curvature radius of the AdS$_5$ 
(we shall set $R=1$ in the following). 
The horizon is located at $r=0$.

In this paper, we focus on adding $N_{\!f}$ {\it massless} 
${\cal N}=2$ hypermultiplets to ${\cal N}=4$ SYM,
keeping the theory at zero temperature.
In the dual gravity this is described by 
a zero temperature ``horizon-crossing'' D7-brane embedding in which the 
distance between the D7 branes and the horizon in the $x_8 - x_9$ direction 
vanishes \cite{Karch:2007br}.

The action for the D7-branes is the Dirac-Born-Infeld (DBI) action
\begin{equation}
S_{\rm DBI} = - N_{\!f}\, T_{\rm D7} \int\! d^8 \xi\, 
  \sqrt{-\textrm{det} (g_{ab} + 2\pi\alpha'\, F_{ab})}\,,
\label{dbi}
\end{equation}
where $T_{\rm D7}$ is the D7-brane tension, $\xi_a$ are worldvolume
coordinates, $g_{ab}$ is the induced worldvolume metric and $F_{ab}$
is the worldvolume $U(1)$ gauge field. At the boundary, 
the gauge field couples to
the $U(1)_B$ ``flavor'' current $J^\mu$, where $U(1)_B$ is the 
``baryon number'' subgroup
of the global symmetry group $U(N_f)$ possessed by the ${\cal N}=2$ 
hypermultiplet fields.
(The exact form of the  $U(1)_B$ current operator is given in 
\cite{Kobayashi:2006sb}.)
Considering finite ``baryon'' density $\langle J^0\rangle \neq 0$ in the 
boundary gauge theory corresponds 
to turning on 
a non-trivial background worldvolume gauge field $A_0(r)$ in the bulk 
\cite{Kobayashi:2006sb}.
The DBI action becomes\footnote{Throughout this paper, we work in 
a gauge with $A_r = 0$.} 
\begin{equation}
\label{DBI_00}
  S_{\rm DBI} = -  \N \, V_3  \int\! dr\,  r^3 \sqrt{1 - A_0'^2} \,,
\end{equation}
where the factor $2\p\a'$ is absorbed into $A_0$, 
$V_3$ is the spatial volume of the boundary gauge theory, 
and the prefactor $\N = \lambda  N_{\!f} N_c /(2\p)^4$ 
is determined by the gauge/gravity duality dictionary \cite{Kobayashi:2006sb}.

The construction above is specific to the D3--D7 system, 
but we can be more general and consider D$q$ probe
branes whose worldvolume include an AdS$_{p+2}$ factor.
For probe branes corresponding to massless flavors the embedding 
is independent of the internal directions. 
One example with $p=2$ would be the defect D5 
on AdS$_4 \times \textrm{S}^2$
in AdS$_5 \times \textrm{S}^5$ \cite{Karch:2000gx}.  
The DBI action reads
\begin{equation}
\label{DBI_000}
  S_{\rm DBI} = -  \N_q \, V_p  \int\! dr\,  r^p \sqrt{1 - A_0'^2} \,,
\end{equation}
where the normalization now includes the tension of the 
$N_{\!f}$ D$q$-branes \cite{Karch:2007br}.
The solution to the embedding problem is given by 
\begin{equation}
\label{sol}
  A_0' = \frac{d}{\sqrt{ r^{2p} + d^2}}\,,
\end{equation}
where $d \equiv (2 \pi \alpha' \N_q)^{-1}\rho$ 
is proportional to the baryon number density $\rho$ \cite{Karch:2007br}. 
In all subsequent formulas, the 
results for the  D3--D7 case are trivially recovered by setting $p=3$ and
using the relation $\alpha'^{-1} = \sqrt\lambda$. 
%
%

\emph{Low-temperature limit of the specific heat.}---
One interesting hint to the nature of the phase of matter described 
by the probe
branes with finite chemical potential is the behavior of 
the specific heat at low temperature. To extract this
information one first needs to generalize the setup to a
 background that contains a black hole,
that is the AdS$_{p+2}$ part of the metric is modified to
\begin{equation}
  ds^2 = \frac{r^2}{R^2} \left[ 
  - \left(  1 - \frac{r_{\!H}^{p+1}}{r^{p+1} }\right) 
 dt^2 + d \vec{x}^2\right]
 + \left(  1 - \frac{r_{\!H}^{p+1}}{r^{p+1} } \right)^{-1}
 \frac{R^2}{ r^2} dr^2\,.
\label{thermal_metric}
\end{equation}
%
Here  $r_{\!H}^{\phantom 1}$ is the horizon radius 
of the black hole related to the temperature of
the black hole by $r_{\!H}^{\phantom 1} = 4 \pi T/(p+1)$ (with $R=1$). 
The action describing the system at finite density 
and finite temperature is almost identical
 to the one in the zero temperature case
\begin{equation}
  S_{\rm \, DBI} = - \N_q \, V_p \int_{r_{\!H}^{\phantom 1}}^\infty \! dr \,
  r^p \sqrt{ 1 - A_0'^2}\,,
\label{dbi77}
\end{equation}
the only difference being that the integration starts 
at $r_{\!H}^{\phantom 1}$ rather than at 0. All
powers of the redshift factor cancel  between $g^{tt}$ and $g^{rr}$. 
The solution for $A_0'$ is still given by Eq.~(\ref{sol}).

The on-shell value of the total 
action~\footnote{The total action is the sum of the 
DBI action (\ref{dbi77}) describing fundamental degrees of freedom 
and the 
bulk gravitational action dual to the adjoint sector of the system.} 
directly gives us minus 
the thermodynamic potential $\Omega$ in the grand canonical ensemble. 
Substituting the solution (\ref{sol}) for $A_0'$  into the action, 
we can write 
$\Omega =  \Omega_{\rm \, ad} + \Omega_{\rm \, fun}$, 
where  $\Omega_{\rm \, ad} \sim T^{p+1}$ is the 
contribution of the adjoint degrees of freedom 
(for the D3/D7 system $\Omega_{\rm \, ad} = - \pi^2 N_c^2 T^4/8$ is 
the free energy of ${\cal N}=4$ SYM at strong coupling), and 
\begin{equation}
\label{freeenergy}
 \Omega_{\rm \, fun}  =  {\cal N}_q \, V_p 
  \int_{r_{\!H}^{\phantom 1}}^\Lambda \! dr \,  
  \frac{r^{2p}}{\sqrt{r^{2p} + d^2}} - \frac{{\cal N}_q}{p+1}\, 
  \int\! d^{p+1} x\, \sqrt{-h (\Lambda)}\, \,,
\end{equation}
%
where $\Lambda$ is the ultraviolet cutoff. In Eq.~(\ref{freeenergy}), 
the local counterterm action built from the metric $h_{\mu\nu}$ induced  
on the slice $r=\Lambda$ by the ambient metric  
(\ref{thermal_metric}) has been added
in the spirit of the holographic renormalization \cite{Skenderis:2002wp}.
%
%
In the grand canonical ensemble, the potential $\Omega$ is a function 
of $T$ and $\mu$, where
$\mu$ is the baryon number chemical potential related to the density 
and temperature 
via the condition 
\begin{equation}
\label{mudrel}
\mu = \int_{r_{\!H}^{\phantom 1}}^{\infty} \! dr \,A_0' \,.
\end{equation}

The integrals in Eqs.~(\ref{freeenergy}) and (\ref{mudrel}) can be expressed
in terms of the Gauss hypergeometric function:
\begin{align}
\Omega_{\rm \, fun} &= \Omega_0 - \frac{  {\cal N}_q  V_p  
  r_{\!H}^{2p+1}}{(2p+1)d} \, 
 \ofo \left( \frac{1}{2}\,, 1+ \frac{1}{2p};
 2 +  \frac{1}{2p}; - \frac{r_{\!H}^{2p}}{d^2} \right)
  + \frac{ {\cal N}_q \, V_p \, r_{\!H}^{p+1}}{2(p+1)} 
  \label{Omega_low_T} \,,\\
\mu &= \mu_0 - r_{\!H}^{\phantom 1} \, 
 \ofo \left( \frac{1}{2}\,,  \frac{1}{2p};
 1 +  \frac{1}{2p}; - \frac{r_{\!H}^{2p}}{d^2} \right)\,,
 \label{mu_low_T}
\end{align}
where $\Omega_0$ and $\mu_0$ are the zero-temperature values,
\begin{equation}\label{muzero}
  \mu_0 = \alpha  \, d^{\frac 1p}\,, \qquad 
  \Omega_0 = - \frac{ {\cal N}_q V_p}{(p+1) \alpha^p}\,  \mu_0^{ p+1 }\,,
\end{equation}
and $\alpha (p) =  \Gamma \left( \frac{1}{2} -  \frac{1}{2p} \right) 
\Gamma \left( 1 +  \frac{1}{2p} \right)/\Gamma \left( \frac{1}{2}\right)$.
We note that the last term in Eq.~(\ref{Omega_low_T}), 
$ {\cal N}_q \, V_p \, r_{\!H}^{p+1}/2(p+1) \equiv 
\Omega_{\rm \, c.t.}$, 
is independent of the matter density and has 
the same temperature dependence as the free energy of adjoint fields 
$\Omega_{\rm \, ad}$. 
We shall focus on the 
density-dependent part of the thermodynamic potential 
$\Delta \Omega \equiv \Omega_{\rm \, fun} - \Omega_{\rm \, c.t.}$.
Equations (\ref{Omega_low_T}) and (\ref{mu_low_T}) determine
$\Delta \Omega$ as a function of temperature and chemical 
potential. At low temperature, both
equations can be treated as series expansions in $T/\mu_0 \ll 1$.
The baryon number density is proportional to $d$,
\begin{equation}
  \rho = - \frac{1}{V_p} \frac{\partial \Omega_{\rm \, fun}}{\partial \mu} 
  = {\cal N}_q\, d\,.  
\end{equation}
One then computes the entropy density $s(\mu, T)$ 
in the grand canonical ensemble 
\begin{equation}
 s (T, \mu)  = \frac{1}{V_p} \, \left( - \frac{\partial \Delta \Omega (T,\mu)}
  {\partial T}\right)_{\mu, V_p} .
\end{equation}
Using Eq.~(\ref{mu_low_T}), we find the entropy density as a function 
of temperature and charge density 
\begin{equation}
s(T,d) = s_0 +  {\cal N}_q  \, \left( \frac{4\pi}{p+1} \right)^{2p+1} \, 
  \frac{T^{2p}}{2d}\,  
\left[ 1 + O \left( T d^{-\frac{1}{p}} \right) \right]\,,
\end{equation}
where $s_0 = 4 \pi \rho/[(p+1) (2 \pi \alpha')]$ is the entropy at zero 
temperature.  This entropy is related to the quark thermal mass (free energy), 
which is negative and proportional to $T$.
Finally, the specific heat (heat capacity per unit volume) 
$c_V^{\phantom 1}$ at constant volume and density 
is determined by
\begin{equation}
c_V^{\phantom 1} 
  = T \left( \frac{\partial s (T,d)}{\partial T}\, \right)_{\rho}\,.
\end{equation}
At low temperature ($T\ll \mu_0$) the density-dependent part of the 
specific heat\footnote{The density-independent part of the specific heat 
is proportional to $T^p$.}
is proportional to $T^{2p}$:
\begin{equation}
\label{specific}
  c_V =  {\cal N}_q \, p  \, \left( \frac{4\pi}{p+1} \right)^{2p+1} \, 
 \frac{T^{2p}}{d}\,  
 \left[ 1 + O \left( T d^{-\frac{1}{p}} \right) \right]\,.
\end{equation}
%
%
This has to be contrasted with
a gas of free bosons whose low temperature specific heat is proportional to
 $T^p$ (a sphere of volume $T^p$ of occupied states in momentum
space, each with energy $T$) or a
gas of fermions, whose low temperature specific heat scales as $T$ for 
any $p$ (a shell of thickness $T$ of occupied states above 
the Fermi surface contributing an energy $T$ each). 
The behavior of the specific heat in Eq.~(\ref{specific}) 
is suggestive of a new type of quantum liquid.

\emph{Zero sound.}---The zero sound mode would manifest itself as 
a pole of the zero-temperature 
retarded flavor current density 
correlator~\cite{LL, Abrikosov, PinesNozieres}. 
In the dual gravity language, the 
pole arises as the quasinormal frequency of the background geometry  
\cite{Birmingham:2001pj,Son:2002sd,Kovtun:2005ev}.
Generically, the quasinormal spectrum is determined by fluctuations 
of all background fields including the metric.
However, in the particular case  we are dealing with, it is sufficient to 
consider fluctuations of the DBI U(1) field in the gravitational 
background (\ref{metric}) with the non-trivial
 background component $A_0$.
Moreover, since the dual quantum field theory is isotropic, 
we can choose the fluctuations to  
depend on time, radial coordinate and one of the spatial coordinates 
(e.g. $x_p$) only
\begin{equation}
\label{fluc}
  A_\mu (r) \rightarrow A_\mu (r) + a_\mu (r,x_0,x_p)\,.
\end{equation}
Substituting (\ref{fluc}) into the DBI action (\ref{DBI_000}) 
and expanding to second order in fluctuations, we find
that the quadratic part of the resulting action is given by 
the sum of the actions describing longitudinal ($a_0$, $a_p$) 
and transverse fluctuations. The action for the longitudinal fluctuations is
(we use the $a_r=0$ gauge)
\begin{equation}
  S^{(2)} = \frac{\N_q}{2} \int\! d^{p+1} x \, dr\, r^p
\Biggl\{ 
\frac{a_0'\,^2}{\left( 1 - A_0'\,^2 \right)^{3/2}} \,  
+
\frac{  \left( \partial_0 a_p - \partial_p a_0\right)^2}
 {r^4 \sqrt{1 - A_0'\,^2}} \,
 -
\frac{  a_p'\,^2 }{\sqrt{1 - A_0'\,^2}}  \Biggr\}\,.
\label{action_long}
\end{equation}
Introducing the Fourier components 
\begin{equation}
a_{\mu} (r, x_0, x_p) = \int\! \frac{d\omega\, d q}{(2\pi)^2} \, 
  e^{-i\omega x_0 + i q x_p} \,
a_{\mu}(r, \omega, q)\,,
\label{fourier}
\end{equation}
we find the equations of motion for the fluctuations
\begin{equation}
\frac{d}{dr}  
 \left[ \frac{r^p\, a_0'}{\left( 1 - A_0'\,^2 \right)^{3/2}} 
    \right] -
\frac{  r^{p-4} }
{ \sqrt{1 - A_0'\,^2}}\,
  \left( \omega q a_p + q^2 a_0 \right)  =0 \,,
\label{eq_1a}
\end{equation}
\begin{equation}
 \frac{d}{dr} \left[ \frac{r^p\,  a_p'}{ \sqrt{1 - A_0'\,^2}}\right] 
+
 \frac{ r^{p-4} }
{ \sqrt{1 - A_0'\,^2}}\, \left( \omega q  a_0 + \omega^2  a_p \right) =0\,.
\label{eq_3a}
\end{equation}
There is also a constraint arising as a consequence of the 
residual gauge invariance of the 
components $a_0$ and $a_p$
\begin{equation}
 \omega  a_0'  + \left( 1  - A_0'\,^2\right) q  a_p' =0\,.
\label{constraint_gen}
\end{equation}
Introducing a new radial coordinate
$z=1/r$, Eqs.~(\ref{eq_1a}, \ref{eq_3a}, \ref{constraint_gen}) 
can be written as
\begin{subequations}\label{fluctuations}
\begin{align}
 & \partial_z \left( f^3 \, z^{2-p} \, a_0' \right) -
f \, z^{2-p}  \left( \omega q  a_p + q^2  a_0 \right) =0\,,\label{gl1}\\
 & \partial_z \left( f^{\phantom 1}\, z^{2-p} \, a_p' \right) + f  \, z^{2-p}  
\left( \omega q  a_0 + \omega^2  a_p \right) =0\,,
\label{gl2}\\
 &  f^2 \, \omega  a_0' +  q  a_p' = 0\,, \label{gl3}
\end{align}
\end{subequations}
where $f(z)=\sqrt{1 + d^2 \, z^{2p}}$.
Following the approach of \cite{Kovtun:2005ev}, 
we use Eq.~(\ref{fluctuations})
to derive  the equation for the 
gauge-invariant variable $E= \omega\, a_p + q \, a_0$:
\begin{equation}
E'' + \left[  \frac{f'\, (3 q^2 - \omega^2 f^2)}
  {f\, (q^2 - \omega^2 f^2)} - \frac{p-2}{z}\right] \, E' + 
\frac{  \omega^2 f^2 - q^2}{f^2}\, E =0\,.
\label{gieq}
\end{equation}
The horizon, $z=\infty$, is an {\it irregular} singular point 
of the differential equation (\ref{gieq}). 
The solution in the vicinity of  $z=\infty$ is given by 
$E(z) \sim  e^{\pm i \omega z}/z$.
The incoming wave boundary condition
  at  the horizon \cite{Son:2002sd} singles out one of the exponents
\begin{equation}
E(z) = C\,  \frac{e^{i \omega z}}{z}\, 
  \Bigl( 1 + O \left( 1/z \right) \Bigr)\,,
\end{equation}
where $C$ is a constant.
For $\omega z \ll 1$ we have
\begin{equation}
E(z) = \frac{C}{z} + i\omega C\,.
\label{match}
\end{equation}
On the other hand, for $\omega z \ll 1$ and $q z \ll 1$ 
with $\omega/q$ fixed, Eq.~(\ref{gieq}) reduces to 
the equation 
\begin{equation}
E'' + \left[  \frac{f'\, (3 q^2 - 
\omega^2 f^2)}{f\, (q^2 - \omega^2 f^2)} - \frac{p-2}{z}\right] \, E'  =0\,,
\label{gieqr}
\end{equation}
whose solution is given in terms of the Gauss hypergeometric function
\begin{equation}
E(z) = C_1 + C_2  z^{p-1} \left[    \frac{q^2}{p f(z)}
  + \frac{(q^2 - p \omega^2)}{p(p-1)} 
\,  \ofo \left(   \frac{1}{2},      \frac{1}{2} -\frac{1}{2p} ; 
 \frac{3}{2} -\frac{1}{2p} ; - d^2 z^{2p} \right) \right]. 
\end{equation}
%
For  $z\rightarrow \infty$ we find
\begin{equation}
E(z) \rightarrow C_1 + C_2 \left( \frac{a}{z} + b \right) +  O(1/z^2)\,,
\end{equation}
where the coefficients $a$ and $b$ are given by 
\begin{equation}
a= \frac{\omega^2}{d}\,, \qquad 
b = \frac{( q^2 - p \omega^2) \, d^{\frac{1}{p}-1} \, 
  \Gamma \left( \frac{1}{2} -  \frac{1}{2p}\right) 
\Gamma \left( \frac{1}{2p}\right) }
{ 2 p^2  \Gamma \left(\frac{1}{2}\right)}\,.
\end{equation}
%
Matching to the expansion (\ref{match}), we find the coefficients $C_1$ and $C_2$:
\begin{equation}
C_1 = \left( i \omega - \frac{b}{a} \right)\, C\,, \qquad C_2 = \frac{C}{a}\,.
\end{equation}
%
The lowest quasinormal frequency is found  by imposing the Dirichlet condition
 at the boundary, $E(0) = 0$ \cite{Kovtun:2005ev}. This condition gives the equation $C_1 =0$ or, equivalently,
 \begin{equation} 
i \, \omega = \left( \frac{q^2}{\omega^2} - p\right) \frac{d^{\frac{1}{p}} 
 \Gamma \left( \frac{1}{2}- \frac{1}{2 p}\right) \Gamma \left( \frac{1}{2 p}\right)}{2p^2\Gamma(\frac{1}{2})}\,.
\label{Dirichlet}
\end{equation}
%
Solving  Eq.~(\ref{Dirichlet}) for small $\omega$ and $q$, we find the dispersion relation for the zero sound
 \begin{equation} 
\label{disp}
\omega = \pm \frac{q}{\sqrt{p}} -  
\frac{i  \Gamma \left(\frac{1}{2}\right) \, q^2}{ d^{\frac{1}{p}}  \Gamma \left( \frac{1}{2}- \frac{1}{2 p}\right) 
\Gamma \left( \frac{1}{2 p}\right)} + O(q^3)\,.
\end{equation}
%
%
Using the expression for the chemical potential at zero temperature 
from Eq.~(\ref{muzero}),
the zero sound dispersion relation can be written as 
 \begin{equation} 
\label{dispmu}
\omega = \pm \frac{q}{\sqrt{p}} - \frac{i\, q^2}{2 p \mu_0}  + O(q^3)\,.
\end{equation}

What is the nature of this excitation?  First, one can exclude the
possibility that it is a superfluid phonon.  Indeed, our background
does not break the particle number symmetry, hence the ground state is
not a superfluid.  Furthermore, superfluid phonon width has a
low-momentum behavior different from $q^2$, namely $q^5$ in 3
spatial dimensions~\cite{LL} and $q^{p+2}$ in $p$ spatial dimensions
(provided that phonon decay is kinematically allowed).
The $q^2$ behavior of the imaginary part is characteristic of zero
sound quantum attenuation, thus we call this mode the zero sound.
Yet in other respects (such as the specific heat temperature dependence)
the system 
does not show Fermi-liquid behavior.
  It is notable
that the zero sound velocity in our system coincides with the velocity
of the finite-temperature first sound, while in a weakly-coupled Fermi liquid 
it is $\sqrt p$ times larger than the first-sound velocity.

%
%



\emph{Conclusion.}---In this paper we have considered a general theory
described by a DBI action in AdS space.  We found that by turning on a
chemical potential one arrives to a new type of quantum liquid.  
The specific heat $c_V^{\phantom 1}$ has an unusual  non-Fermi liquid 
$T^{2p}$ behavior ($T^6$ in 3+1 dimensions and $T^4$ in 2+1
dimensions).  The low energy spectrum contains a gapless mode with a
dispersion relation similar to the zero sound in Fermi liquids.  One
can speculate that the mode observed here is what the Fermi-liquid
zero sound becomes when the interaction is infinitely strong.  In this
connection, we note that in a simple model of the Fermi liquid, the
velocity of the zero and first sounds approach each other in the limit
where the interaction strength (parameterized by the Fermi-liquid
parameter $F_0$) is infinite~\cite{PinesNozieres}.

The systems described here are strongly coupled, as they have gravity duals.
It would be interesting to investigate the properties of the ground state
and the zero sound in the weak-coupling regime of the ${\cal N}=4$ SYM
theory with ${\cal N}=2$ matter hypermultiplets.  We leave this problem
for future work.

\vspace{4pt}

\emph{Acknowledgments.}---%
The work of A.K. was supported, in part, by  U.S.\ Department
    of Energy under Grant No.~DE-FG02-96ER40956.
The work of D.T.S was supported, in part, by the U.S.\ Department
of Energy under Grant No.~DE-FG02-00ER41132.
The work of A.O.S. was supported, in part, by the UK STFC Advanced Fellowship.




\end{document}